\documentclass[iop,apjl]{emulateapj}

\usepackage[colorlinks,citecolor=blue]{hyperref} 
\usepackage[all]{hypcap} 

\usepackage{graphicx}

\def\equationautorefname~#1\null{Equation~(#1)\null}

\newcommand{\re}{\hbox{$R_{\rm e}$}}
\newcommand{\atl}{ATLAS$^{\rm 3D}$}

\slugcomment{The Astrophysical Journal Letters. Received 2015 January 15; accepted 2015 March 31}
\shorttitle{Nearly-isothermal mass profiles to 4\re\ in ETGs}
\shortauthors{Cappellari et al.}

\begin{document}

\title{Small scatter and nearly-isothermal mass profiles to four half-light radii\\ from two-dimensional stellar dynamics of early-type galaxies}

\author{Michele Cappellari\altaffilmark{1}, 
Aaron J.~Romanowsky\altaffilmark{2,3},  
Jean P.~Brodie\altaffilmark{2}, 
Duncan A.~Forbes\altaffilmark{4},  
Jay Strader\altaffilmark{5}, 
Caroline Foster\altaffilmark{6}, 
Sreeja S.~Kartha\altaffilmark{4},
Nicola Pastorello\altaffilmark{4}, 
Vincenzo Pota\altaffilmark{2,4}, 
Lee R.~Spitler\altaffilmark{6,7}, 
Christopher Usher\altaffilmark{4}, 
Jacob A. Arnold\altaffilmark{2}}

\affil{
\altaffilmark{1}Sub-department of Astrophysics, Department of Physics, University of Oxford, Denys Wilkinson Building, Keble Road, Oxford, OX1~3RH, UK\\
\altaffilmark{2}University of California Observatories, 1156 High Street, Santa Cruz, CA 95064, USA\\
\altaffilmark{3}Department of Physics and Astronomy, San Jos\'e State University, One Washington Square, San Jose, CA 95192, USA\\
\altaffilmark{4}Centre for Astrophysics \& Supercomputing, Swinburne University, Hawthorn, VIC 3122, Australia\\
\altaffilmark{5}Department of Physics and Astronomy, Michigan State University, East Lansing, Michigan 48824, USA\\
\altaffilmark{6}Australian Astronomical Observatory, PO Box 915, North Ryde, NSW 1670, Australia\\
\altaffilmark{7}Department of Physics and Astronomy, Macquarie University, North Ryde, NSW 2109, Australia
}

\begin{abstract}
We study the total mass-density profile for a sample of 14 fast-rotator early-type galaxies (stellar masses $10.2\la\log M_\ast/M_\odot\la11.7$). We combine observations from the SLUGGS and \atl\ surveys to map out the stellar kinematics in two-dimensions, out to a median radius for the sample of four half-light radii \re\ (or 10 kpc), and a maximum radius of 2.0--6.2 \re\ (or 4--21 kpc). We use axisymmetric dynamical models based on the Jeans equations, which allow for a spatially varying anisotropy, and employ quite general profiles for the dark halos, and in particular do not place any restriction on the profile slope. This is made possible by the availability of spatially extended two-dimensional kinematics. We find that our relatively simple models provide a remarkably good description of the observed kinematics. The resulting {\em total} density profiles are well described by a nearly-isothermal power law $\rho_{\rm tot}(r)\propto r^{-\gamma}$ from \re/10 to at least 4\re, the largest average deviation being 11\%. The average logarithmic slope is $\langle\gamma\rangle=2.19\pm0.03$ with {\em observed} rms scatter of just $\sigma_\gamma=0.11$. This scatter out to large radii, where dark matter dominates, is as small as previously reported by lensing studies around $r\approx\re/2$, where the stars dominate. Our bulge-halo conspiracy places much tighter constraints on galaxy formation models. It illustrates the power of two-dimensional stellar kinematics observations at large radii. It would now be important to test the generality of our results for different galaxy types and larger samples.
\end{abstract}


\section{Introduction}

\begin{table*}
\centering
\caption{Sample of fast rotator early-type galaxies and measured parameters}
\tabcolsep=3.6pt
\begin{tabular}{cccccccccccc}
\tableline
Name        & $D$        & $\log\re$     & $R_{\rm max}$  & $R_{\rm max}/\re$ & $\langle\gamma\rangle_r$ & $\langle\gamma\rangle_r$ & $\langle\gamma\rangle_r$ & $\langle|\Delta|\rangle_r$ & MGE ref. & $\chi_{\rm JAM}^2/\chi_{\rm LOESS}^2$ & ${\rm median}(|\delta V_{\rm rms}|)$ \\
            & (Mpc)    & (\arcsec)     & (kpc)         &                   & (0.1--1\re)            & (1--4\re)              & (0.1--4\re)            & (0.1--4\re)            &         &       & (\%) \\
(1)         & (2)      & (3)           & (4)           & (5)               & (6)                    & (7)                    & (8)                    & (9)                    & (10)    & (11)  & (12)\\
\tableline
NGC 0821    &  23.4    &      1.60     &     13.2      &              2.92 &      -2.16             &   -2.36                &    -2.23               &        0.016           &   C06   & 1.09  & 8.5 \\
NGC 1023    &  11.1    &      1.68     &     11.3      &              4.39 &      -2.20             &   -2.19                &    -2.20               &        0.009           &   S09   & 0.98  & 12.3 \\
NGC 2768    &  21.8    &      1.80     &     19.7      &              2.95 &      -2.10             &   -1.82                &    -2.01               &        0.025           &   C06   & 1.02  & 5.1 \\
NGC 2974    &  20.9    &      1.58     &     10.8      &              2.80 &      -2.22             &   -2.49                &    -2.30               &        0.015           &   C06   & 1.10  & 14.9 \\
NGC 3115    &   9.5    &      1.54     &     10.1      &              6.19 &      -2.28             &   -2.53                &    -2.37               &        0.028           &   E99   & 1.10  & 10.7 \\
NGC 3377    &  10.9    &      1.55     &      8.7      &              4.64 &      -2.22             &   -1.75                &    -2.05               &        0.046           &   C06   & 0.99  & 8.4 \\
NGC 4111    &  14.6    &      1.08     &      3.8      &              4.46 &      -2.12             &   -2.16                &    -2.13               &        0.043           &   ---   & 0.85  & 7.7 \\
NGC 4278    &  15.6    &      1.50     &      9.0      &              3.76 &      -2.19             &   -2.45                &    -2.29               &        0.028           &   C06   & 1.09  & 7.8 \\
NGC 4473    &  15.3    &      1.43     &      8.4      &              4.18 &      -2.12             &   -2.29                &    -2.18               &        0.015           &   C06   & 1.20  & 6.4 \\
NGC 4494    &  16.6    &      1.69     &      8.0      &              2.04 &      -2.18             &   -2.52                &    -2.26               &        0.045           &   S13   & 1.06  & 9.2 \\
NGC 4526    &  16.4    &      1.65     &     18.2      &              5.13 &      -2.21             &   -2.29                &    -2.24               &        0.014           &   C06   & 0.97  & 9.0 \\
NGC 4649    &  17.3    &      1.82     &     21.4      &              3.86 &      -2.10             &   -2.34                &    -2.19               &        0.020           &   S13   & 1.05  & 8.9 \\
NGC 4697    &  11.4    &      1.79     &     12.5      &              3.66 &      -2.16             &   -2.34                &    -2.23               &        0.023           &   S13   & 0.99  & 7.1 \\
NGC 7457    &  12.9    &      1.56     &      6.9      &              3.05 &      -1.82             &   -2.23                &    -1.96               &        0.036           &   C06   & 1.33  & 6.1 \\
\tableline
\end{tabular}
\tablecomments{
Column~(1): galaxy name; 
Columns~(2)-(3): distance and half-light radius from \citet{Cappellari2011a}; 
Columns~(4)-(5): Maximum radius $R_{\rm max}$ sampled by the stellar kinematics; 
Column~(6): Average logarithmic slope $\langle\gamma\rangle_r=\Delta\log\rho_{\rm tot}/\Delta\log r$ of the total mass profile in the interval $0.1\re<r<\re$; 
Column~(7): as in Col.~6 for $\re<r<\min(4\re,R_{\rm max})$;
Column~(8): as in Col.~6 for $0.1\re<r<\min(4\re,R_{\rm max})$;
Column~(9): Average absolute deviation between $\rho_{\rm tot}$ and the best-fitting power-law (in dex);
Column~(10): Reference for the MGE model: C06$=$\citet{Cappellari2006}, S09$=$\citet{Scott2009}, S13$=$\citet{Scott2013p21}, E99$=$\cite{Emsellem1999}
Column~(11): Quality of fit, where $\chi_{\rm JAM}$ is measured from the JAM models and $\chi_{\rm LOESS}$ from the smoothed data. In both cases the reference is the original SLUGGS data, excluding values with zero final weight. This ratio approximates $\chi^2/{\rm DOF}$, but is insensitive to the normalization of the kinematic uncertainties.
Column~(12): Median of the absolute relative deviations $|\delta V_{\rm rms}|\equiv |V_{\rm rms}/V_{\rm rms}^{\rm LOESS}-1|$ between the original and the smoothed SLUGGS $V_{\rm rms}$.
}
\label{tab1}
\end{table*}

One of the pillars on which our entire paradigm of galaxy formation rests is the fact that dark matter of an unknown nature dominates the mass budget of the Universe \citep{Blumenthal1984}. Without dark matter, the primordial gas would not be able to collapse sufficiently quickly within the center of dark matter halos to form the galaxies we observe \citep{White1978}.

Total mass-density profiles of spiral galaxies were obtained decades ago, given the simple geometry of their spiral disks and the availability of ionized \citep{Rubin1980} and neutral gas \citep{Bosma1978}, which provides a kinematical tracer easy to measure and model (see review by \citealt{Courteau2014}). The measurements indicated flat circular rotation curves and consequently nearly-isothermal $\rho_{\rm tot}\propto r^{-2}$ profiles.

The situation is very different for early-type galaxies (ETGs: ellipticals and lenticulars), which by definition lack well-defined spiral disks and are cold-gas poor. For ETGs one usually has to rely on expensive observations of the stellar kinematics and more complex dynamical models. This has restricted most studies to radii not much larger than the half-light radius \re.

A general consensus has emerged for the mass distribution of the ETGs inner parts. Long-slit observations of two different samples of $\approx20$ ETGs revealed rotation curves to be nearly flat with nearly-isothermal mass profiles, as in spiral galaxies, within the median radius $r\approx2\re$ sampled by the kinematics \citep{Gerhard2001,Thomas2011}. This agrees with strong gravitational lensing studies finding nearly-isothermal slopes, with small scatter, for the {\em total} galaxy density profile of 73 ETGs, at a typical radius of $r\approx\re/2$  \citep{Auger2010}.
These central slopes are similar to those measured for group/cluster-scale profiles using X-ray modeling \citep[e.g.][]{Humphrey2010} or stacked weak-lensing technique \citep[e.g.][]{Gavazzi2007}.

The largest detailed study of dark matter in galaxy centers was based on the integral-field observations of the volume-limited \atl\ sample of 260 ETGs \citep{Cappellari2011a}. It inferred a median dark matter fraction of just $f_{\rm DM}(\re)=13\%$, within a sphere of radius \re, over the full sample \citep{Cappellari2013p15}. {\em This shows that studies restricted to the central regions of ETGs mainly measure the stellar mass distribution.}

To explore the region where dark matter dominates, one needs to reach at least $\sim4\re$. Existing studies at these radii targeted one galaxy at a time. They used observations of extended HI disks \citep{Weijmans2008}, deep stellar kinematics at a few sparse locations \citep{Weijmans2009,Forestell2010,Murphy2011}, globular cluster kinematics \citep{Napolitano2014} and planetary nebulae \citep{Romanowsky2003,deLorenzi2008,deLorenzi2009,Das2011,Napolitano2011,Morganti2013}. 

The fact that galaxies were modeled using different techniques or kinematic tracers, and for a mix of ETGs with axisymmetric and triaxial shapes, resulted in a still unclear picture of the global trends. Tentative conclusions suggest a trend with massive ETGs having nearly isothermal total mass profiles and flat circular velocities  \citep[see review by][]{Gerhard2013}, while lower mass ones have more slowly rising dark matter profiles and correspondingly falling circular velocity profiles \citep[also][]{Romanowsky2003}.

The situation is changing with the availability of extended stellar kinematics for significant samples of ETGs \citep{Brodie2014,Raskutti2014}. Here we present the first detailed and fully homogeneous analysis using larges-scale stellar dynamics of a statistically significant sample of 14 ETGs (but see \citealt{Deason2012dm} for less detailed models). The key novelty of this work is the availability of {\em two-dimensional} stellar kinematics out to a median radius for the sample of $r\approx4\re$ from the SLUGGS survey \citep{Brodie2014}, which we combine with integral-field kinematics in the central regions ($\sim1\re$) from the \atl\ survey \citep{Cappellari2011a}.

\begin{figure*}
\centering
\includegraphics[width=0.35\textwidth]{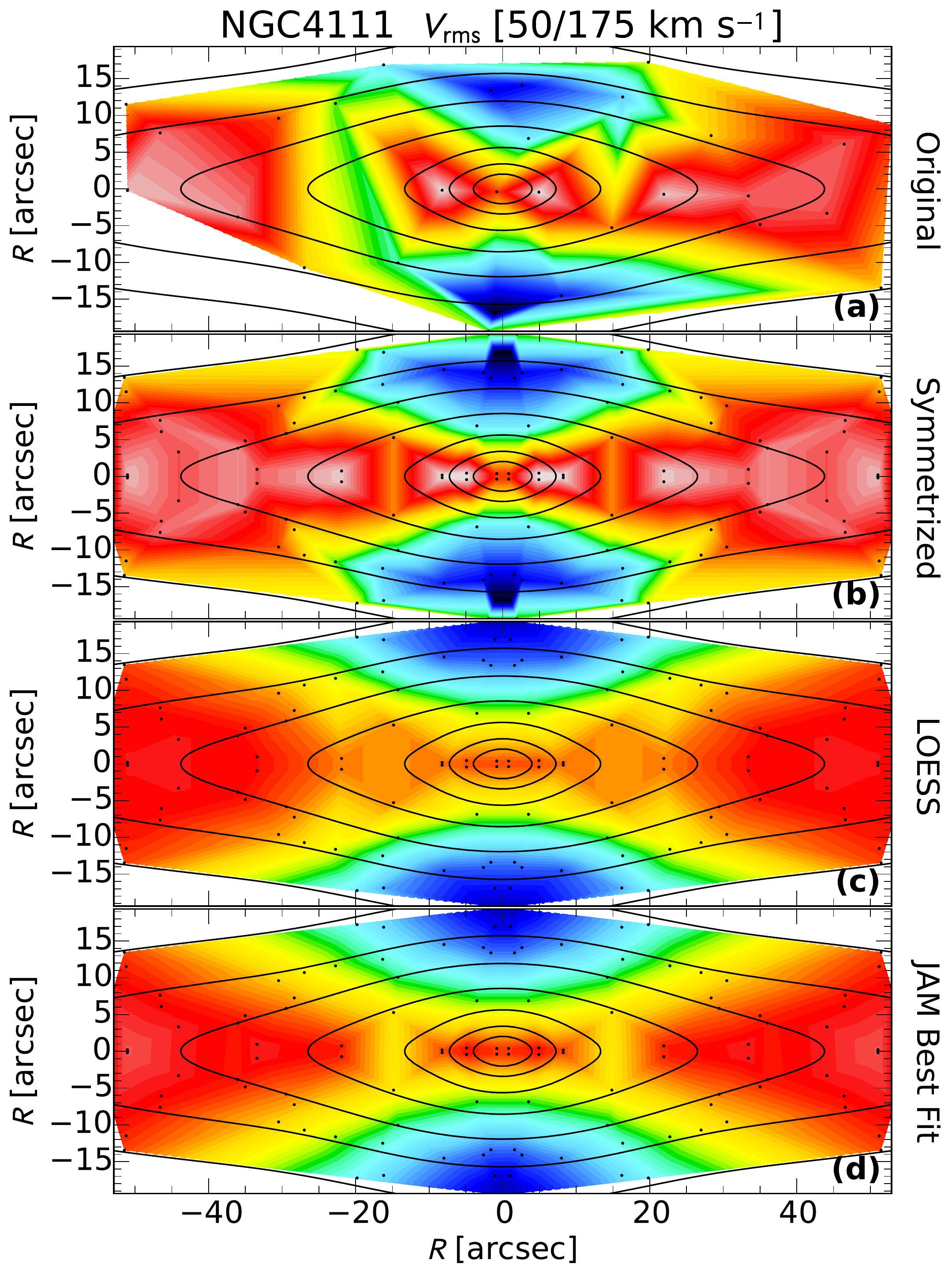}
\includegraphics[width=0.35\textwidth]{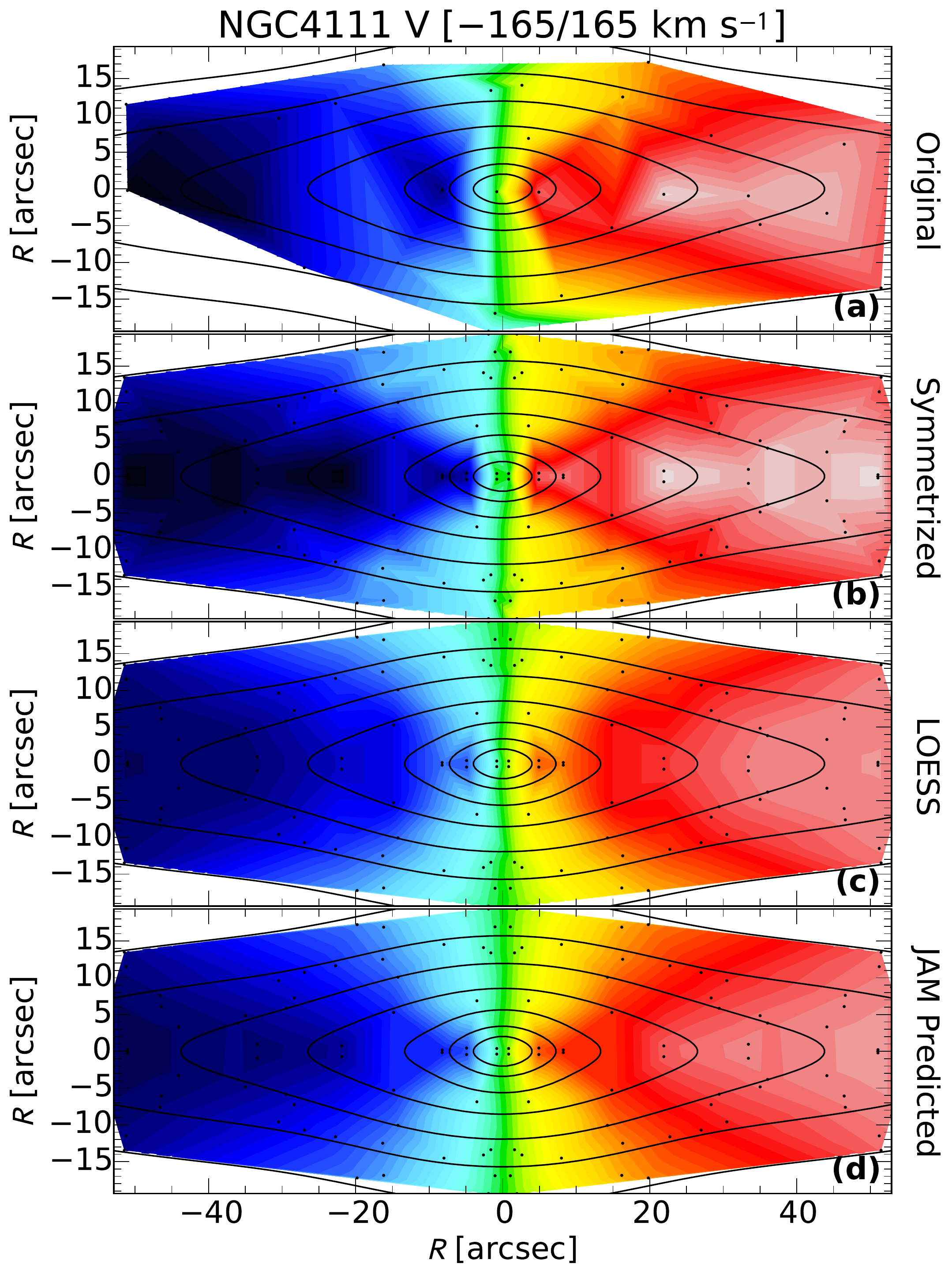}
\caption{Symmetrization and cleaning of SLUGGS data. Left panels show the stellar $V_{\rm rms}\equiv\sqrt{V^2+\sigma^2}$, while right ones show the mean stellar velocity $V$. (a) original linearly-interpolated data; (b) bi-symmetrized data; (c) bi-symmetrized and LOESS-smoothed data. Note the good prediction, in the central parts, of the actual SAURON kinematics in \autoref{fig:jam_fits1}; (d) best-fitting JAM model. The right-panel shows the JAM {\em prediction} for the mean velocity using the same model parameters as in the left-panel, adopting for all Gaussians $\sigma_\phi=\sigma_R$ and $\kappa=1.20$. We tried different $\sigma_\phi/\sigma_R$ ratios$^3$ but found best fits for $\sigma_\phi\approx\sigma_R$ (see \citealt{Cappellari2008} for definitions).\label{fig:symm}}
\end{figure*}

\section{Sample and Data}

\subsection{Axisymmetric sample selection}

We study a subsample of the SLUGGS ETG sample, with kinematics from \citet{Arnold2014}. We want a homogeneous set of galaxies and we use an axisymmetric dynamical modelling method. For this reason we only select ETGs classified as fast rotator in \citet{Emsellem2011}. We add the fast rotator NGC~3115, without \atl\ kinematics. Our sample of 14 galaxies is given in \autoref{tab1} (stellar masses $10.2\la\log M_\ast/M_\odot\la11.7$ from \citealt{Cappellari2013p15}). 
  
Fast rotators were shown by \citet{Krajnovic2011} to have kinematical axes, within about 1\re, extremely well aligned with the {\em photometric} major axes measured at the much larger radii $\sim3\re$ sampled by the SLUGGS kinematics. This alignment, for the statistically-significant, \atl\ sample, shows that fast rotator ETGs, as a class, must be axisymmetric out to the region sampled by our models. Significant deviations from axisymmetry would produce a broad distribution of kinematical misalignment that is clearly ruled out by the \atl\ data. The SLUGGS kinematics of fast rotators is also generally consistent with axisymmetry and shows small kinematical misalignment. 
 
Some genuine deviations from a bi-symmetric velocity field do exist in the SLUGGS data. Notable examples are NGC~4494 in the maps of \citet{Arnold2014} and NGC~4473 in \citet{Foster2013}. However, these deviations are more likely due to unrelaxed substructure in the stellar halo, than to non-axisymmetry in the relaxed stellar distribution. In fact, the tight alignment between \atl\ stellar kinematics and photometry, as well as the complete lack of photometric twist, is extremely unlikely in a triaxial configuration \citep[e.g.][]{Binney1985}. In this work we attribute deviations from the axisymmetric assumption to either sub-structure or systematic problems in the data, and simply try to remove their effects from our models. 

\subsection{Photometry and mass models}

We use the Multi-Gaussian Expansion (MGE) \citep{Emsellem1994,Cappellari2002mge} to parametrize both stellar and dark halo distributions. The MGEs of 8 galaxies come from \citet{Cappellari2006} and one from \citet{Scott2009}. They include deep $I$-band photometry out to 5--10\re. Three SDSS $r$-band MGEs come from \citet{Scott2013p21}. The $V$-band  MGE for NGC~3115 comes from \citet{Emsellem1999}. The MGE for NGC4111 was redone on the $i$-band SDSS image using the Python version\footnote{Available from \url{http://purl.org/cappellari/software}} of the MGE-fitting method of \citet{Cappellari2002mge}, to avoid the $r$-band saturation \citep[see][supplementary material]{Scott2013p21}.

\subsection{Stellar kinematics}

The kinematics\footnote{Available from \url{http://purl.org/atlas3d}} for the inner ($\sim1\re$) region of our galaxies was observed with the SAURON integral-field spectrograph and was homogeneously extracted for the \atl\ survey as described in \citet{Cappellari2011a}. For 10/14 galaxies the observations, taken as part of the SAURON survey, were already presented in \citet{Emsellem2004}. The kinematics for the outer regions were observed with the Keck/DEIMOS multi-slit spectrograph by the SLUGGS survey. They were presented and described in \citet{Arnold2014}.

Briefly, in both cases the kinematics were extracted in pixel space using the pPXF method \citep{Cappellari2004}. The \atl\ data covered the H$\beta$ and Mg~$b$ spectral region (4800--5380 \AA) and were extracted using the full MILES stellar library \citep{Sanchez-Blazquez2006} as templates. While the SLUGGS data covered the \ion{Ca}{2} triplet spectral region (8480-8750 \AA) and were extracted using the stellar library of \citet{Cenarro2001}.

\begin{figure*}[p]
\centering
\includegraphics[height=9.5cm]{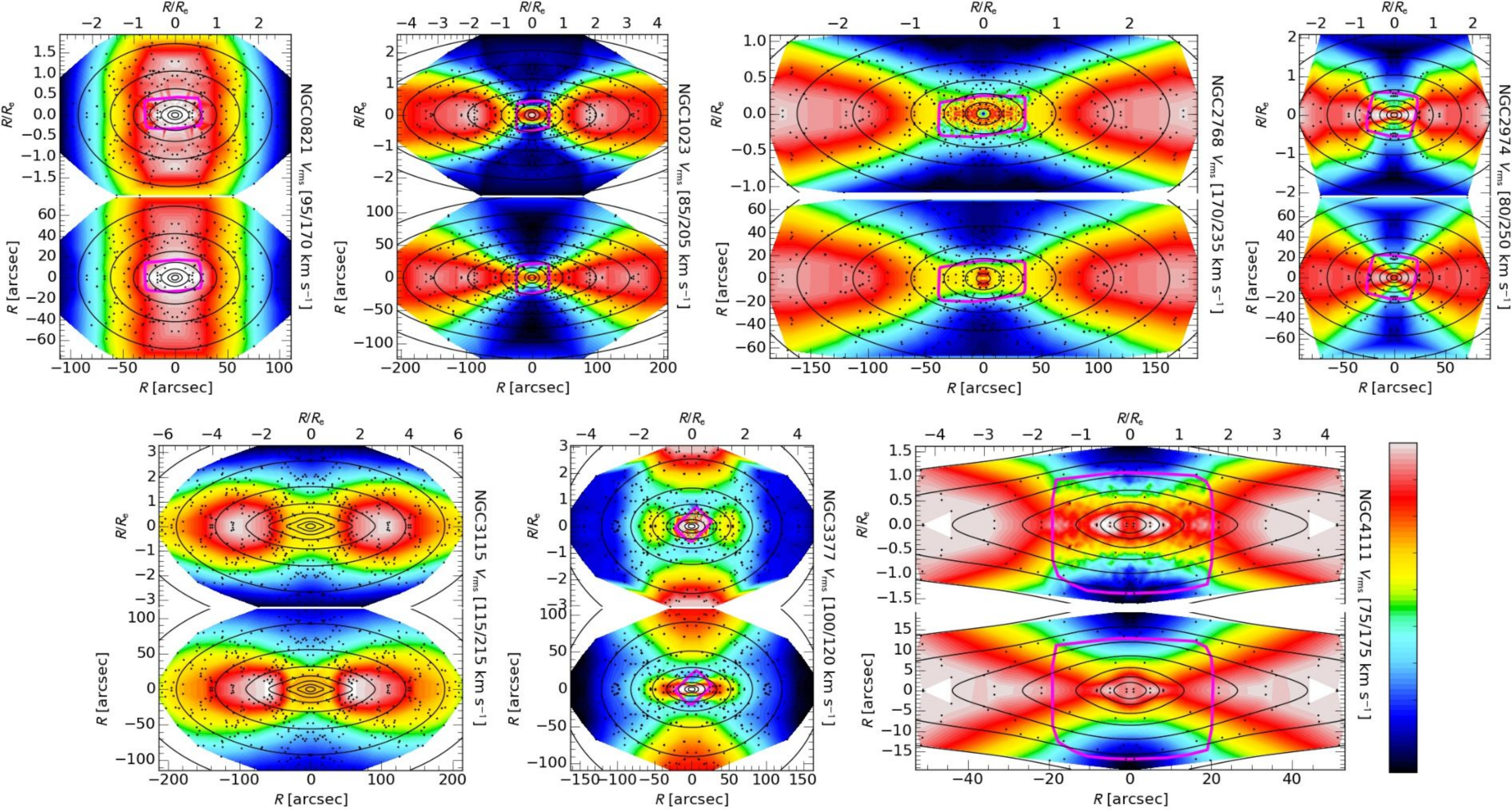}
\caption{Dynamical models from two-dimensional stellar kinematics. In each panel, the top panel shows the observed stellar $V_{\rm rms}$, while the bottom panel shows the JAM model prediction. The spatial scale of the top panels is in units of the half-light radius \re. Contours of the MGE surface brightness in 1 mag steps are overlaid. The data include DEIMOS/SLUGGS stellar kinematics at large radii. They were bi-symmetrized and LOESS smoothed as described in the text (see \autoref{fig:symm}). The locations sampled by the data are shown as black dots. The SAURON/\atl\ data, used at smaller radii, are enclosed by the magenta lines.}
\label{fig:jam_fits1}
\end{figure*}

\begin{figure*}[p]
\centering
\includegraphics[height=9.5cm]{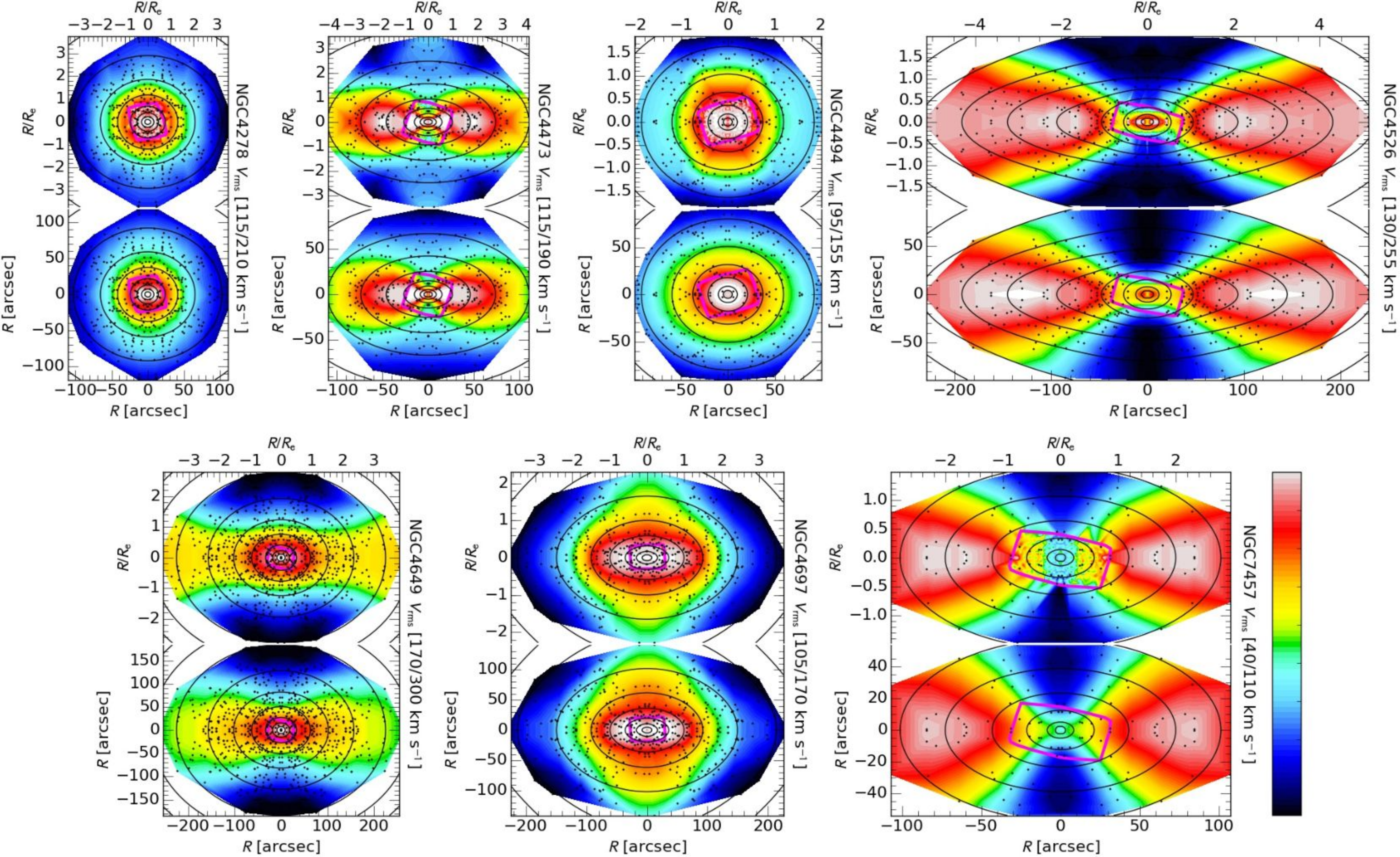}
\caption{Same as in \autoref{fig:jam_fits1} for the second half of the galaxy sample.}
\label{fig:jam_fits2}
\end{figure*}

\section{Methods}

\subsection{Symmetrization and outliers removal}

\footnotetext{This is done by replacing $\mathcal{D}$ in equation~(38) of \citet{Cappellari2008} with $[\mathcal{D}+(1-c_k) b_k q_k^2\sigma_k^2/R^2]$ (Note 9 of the arXiv version of that paper).}

A key step of this analysis is the removal of potential outliers from the SLUGGS kinematics. These may have significant effects on the modeling results, owing to the relatively sparse sampling of the kinematics at large radii. 

When fitting kinematics to models, one can enforce symmetry during the kinematic extraction \citep[e.g.][]{Cappellari2004}. We adopt here a conceptually similar idea, which is easier to apply to irregularly sampled or discrete kinematics (see also \citealt{Morganti2013}).

At every observed location $(x_j,y_j)$ on the sky, where the $x$-axis coincides with the kinematic major axis PA$_{\rm kin}$ (taken from  \citealt{Krajnovic2011}), we calculate the rms velocity $V_{{\rm rms},j}^2\equiv V_j^2 + \sigma_j^2$ from the SLUGGS data, where $V_j$ is the mean stellar velocity and $\sigma_j$ is the velocity dispersion. We then generate a new set of bi-symmetric positions by replicating four times the observed values as follows:
\begin{equation}
\left\{
\begin{array}{rcl}
x_j&\rightarrow&(x_j,-x_j,x_j,-x_j)\\
y_j&\rightarrow&(y_j,y_j,-y_j,-y_j)\\
V_{{\rm rms},j}&\rightarrow&(V_{{\rm rms},j},V_{{\rm rms},j},V_{{\rm rms},j},V_{{\rm rms},j})
\end{array}\right.
\label{eq:symvrms}
\end{equation}
The measurement uncertainties are increased by $2\times$. Given the bi-symmetry of the model, the $\chi^2$ statistics, describing the quality of the fit, is unchanged.
\autoref{eq:symvrms} applies to the even velocity moments like $V_{\rm rms}$ and $\sigma$. While for the odd velocity moments like $V$, the last line is replaced by:
\begin{equation}
V_j\rightarrow(V_j,-V_j,V_j,-V_j).
\label{eq:symvel}
\end{equation}
The resulting kinematics measurements are then smoothed with the (Python version) of the two-dimensional LOESS algorithm of \citet{cleveland1979robust}, implemented$^1$ by \citet{Cappellari2013p20} (see example in \autoref{fig:symm}). 

Comparison between the smoothed/cleaned $V_{\rm rms}^{\rm LOESS}$ and the original one suggests the SLUGGS error being underestimated due to systematics, with typical errors of $\sim10\%$ (\autoref{tab1}). The original errors would indicate non-smooth kinematics, which seems inconsistent with the smooth galaxy images.

In the following models we do not fit the original kinematics, but the symmetrized/smoothed one. This has the added benefit of making the agreement between data and model easy to visually assess. We need to verify that our models capture the global features of the data, and that our results are not driven by a few deviant values. This is important in situations where data systematics may be present and relying entirely on $\chi^2$ statistics may be misleading.

\subsection{Weighting and matching of the two datasets}

Another issue for the modeling is the fact that the \atl\ observations consist of many more data points with smaller uncertainties, which completely dominate the $\chi^2$ estimate.  However here we want our dark halo determinations to be especially constrained by the SLUGGS data, which sample the region where the halo dominates. Similarly to \citet{Morganti2013}, we increase the size of the \atl\ kinematic uncertainties so that, for a good fit, the two datasets provide an equal contribution to the $\chi^2$.  We leave the SLUGGS uncertainties unchanged, to retain properly normalized confidence levels for our model parameters. 

The SLUGGS $V_{\rm rms}$ at the SAURON locations tend to be lower than the measured SAURON values. We find a median offset of 11\%, which is larger than the 5\% level we consider unavoidable between independent datasets. This offset was noted by \citet{Arnold2014}, but its source is unknown. The \atl\ data agree on average with hundreds of independent determinations from the literature (fig.~8 of \citealt{Cappellari2013p15}), suggesting the SLUGGS data may be offset with respect to the optical literature. Here we simply multiply the SLUGGS $V_{\rm rms}$ to fit, for each galaxy, the interpolated SAURON data at the same locations. This is the standard kinematics we fit with our models. However, importantly, we have also run all our models with the SLUGGS data alone, and confirmed that the slopes of the total mass profiles agree with those of our standard models.

\subsection{Dynamical models}

We model the \atl\ and SLUGGS stellar kinematics using the (Python version) of the axisymmetric Jeans anisotropic modeling (JAM) method$^1$ \citep{Cappellari2008}. The approach is the same used in \citet{Cappellari2013p15} for the \atl\ data alone. It employs a Bayesian method with constant (i.e., `ignorant') priors on all parameters.

The key difference between this work and previous stellar dynamical studies of dark halos in ETGs is the fact that we place virtually no constraint on the halo profile parameters. The halo is assumed to be spherical, but is described by a generalized \citet{navarro96} profile (gNFW) with free normalization, inner slope and break radius:
\begin{equation}
    \rho_{\rm DM}(r)=\rho_s\left(\frac{r}{r_s}\right)^\alpha
        \left(\frac{1}{2}+\frac{1}{2}\frac{r}{r_s}\right)^{-\alpha-3}.
\end{equation}

Our models have 7 free parameters. Some are poorly constrained but are not of interest here. They are just `nuisance parameters', marginalized over to derive the total mass profiles studied here. The parameters are: (i) the inclination $i$; (ii) the anisotropy $\beta_z\equiv1-\sigma_z^2/\sigma_R^2$, with $\sigma_z$ and $\sigma_R$ the stellar dispersion in cylindrical coordinates, for the MGE Gaussians with $\sigma_j<\re$; (iii) the anisotropy for the remaining Gaussians at larger radii; (iv) the stellar $(M/L)_{\rm stars}$; (v) the break radius of the dark halo, constrained to be $10<r_s<50$ kpc; (vi) the halo density $\rho_s$ at $r_s$; and (vii) the dark halo slope $\alpha$ for $r\ll r_s$.

\section{Results}

\begin{figure*}
\centering
\includegraphics[width=0.45\textwidth]{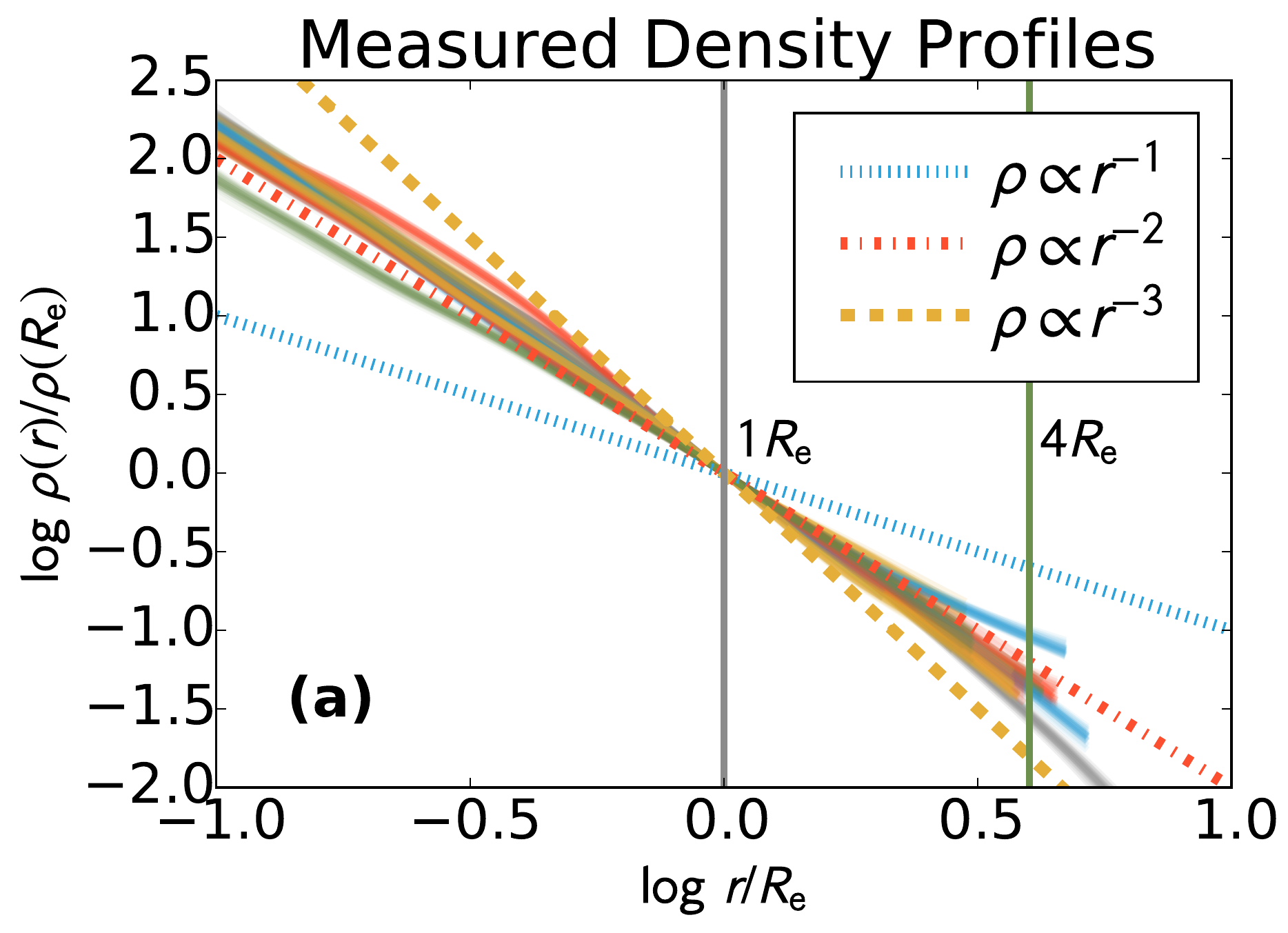}
\includegraphics[width=0.45\textwidth]{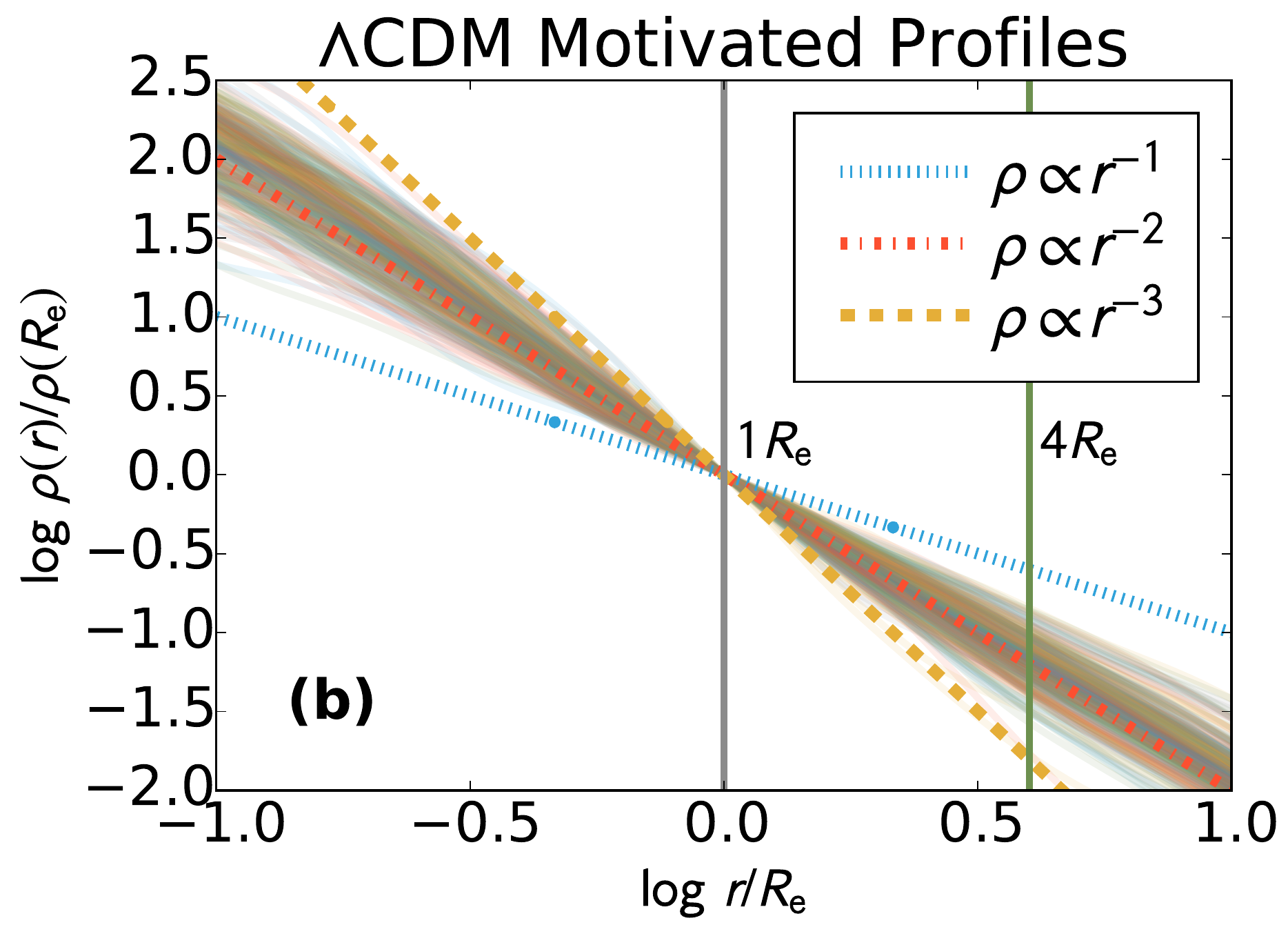}
\includegraphics[width=0.45\textwidth]{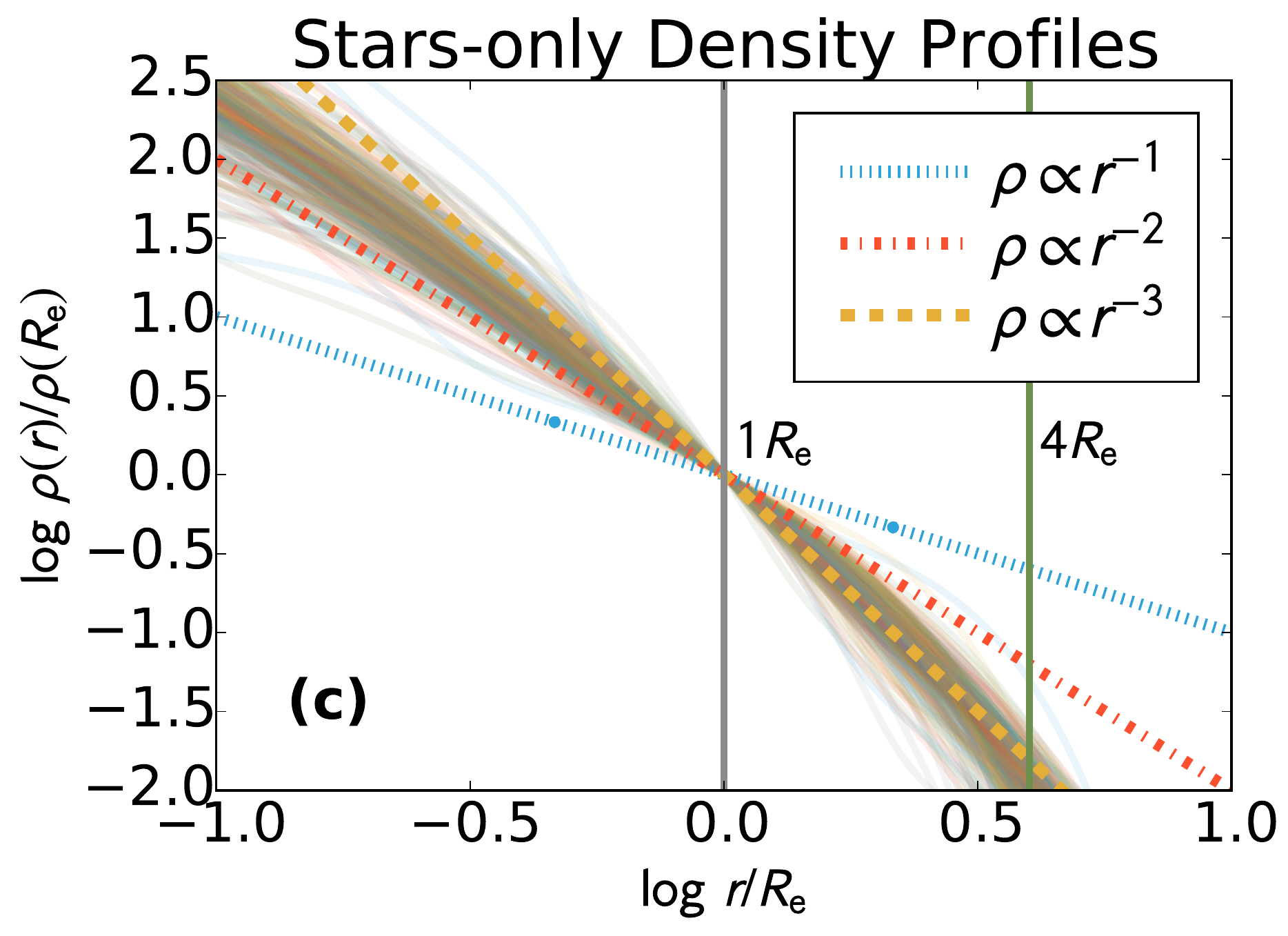}
\includegraphics[width=0.45\textwidth]{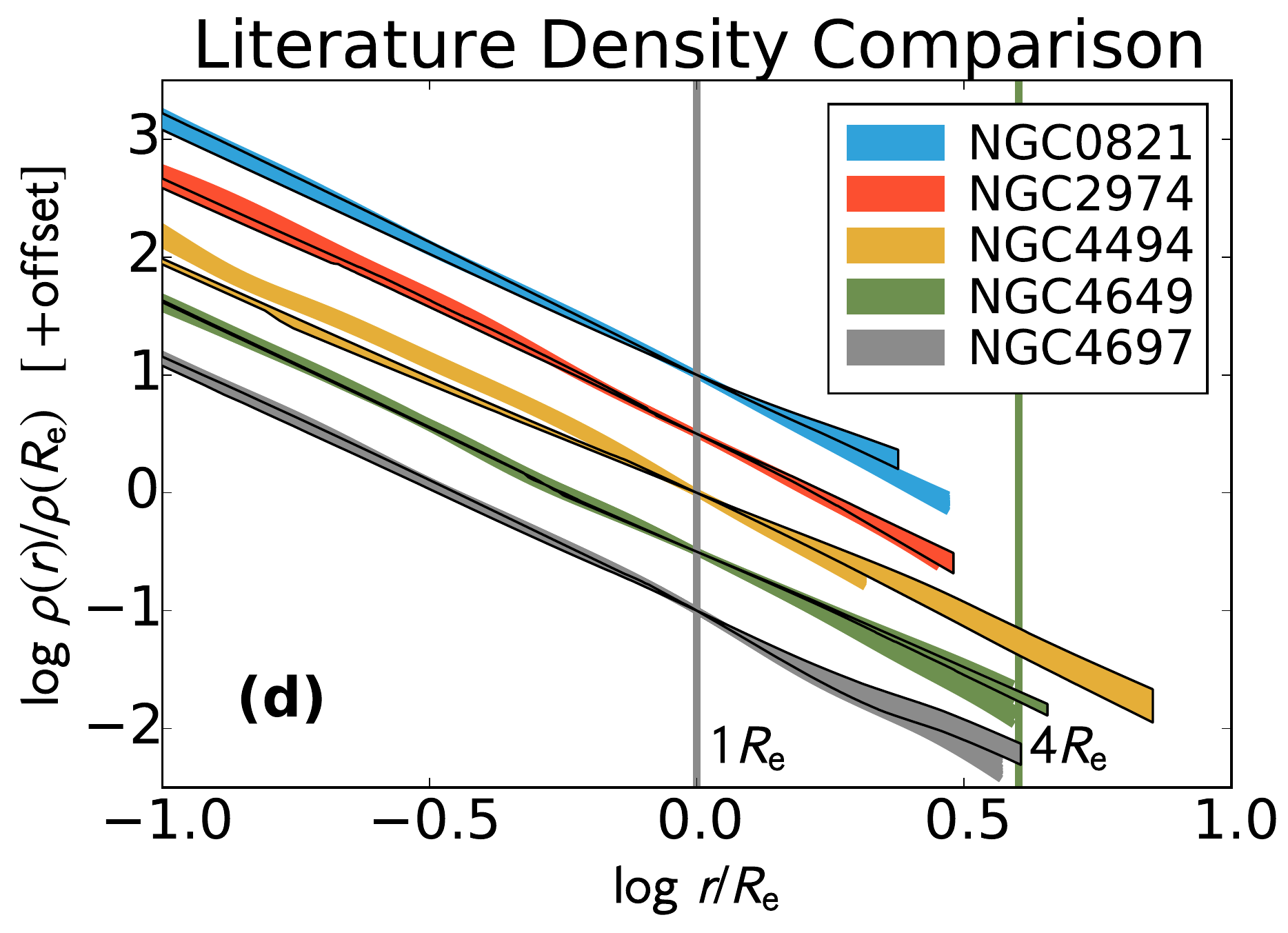}
\caption{Profiles of the total mass-density distribution. (a) Measured profiles for the 14 modeled galaxies with SLUGGS$+$\atl\ data. The profile for each galaxy was plotted for 100 realizations randomly drawn from the posterior distribution of the model parameters, to illustrate the random model uncertainties. Three lines with $\rho\propto r^{-1}$ (NFW inner slope), $\rho\propto r^{-2}$ (isothermal) and $\rho\propto r^{-3}$ are also shown. (b) Cosmologically-motivated profiles (these are models E from \citealt{Cappellari2013p15}). These were computed by attaching, to the stellar density of the \atl\ galaxies, spherical NFW dark halos with masses predicted by the abundance matching technique in such a way that the models fit the stellar kinematics. These models naturally predict a nearly isothermal total-mass profile out to $r\ga10\re$. (c) For comparison the purely-stellar profiles of the same \atl\ galaxies in (b) are shown. (d) Comparison between our density profiles and published ones. The profiles with outline enclose the allowed range of published profiles, while the colored bands are realizations from our models posterior.
}
\label{fig:mass_profiles}
\end{figure*}

\subsection{The models describe the data well}

The first result is the simple fact that the models provide a good description of the stellar kinematics of all the modeled galaxies over the full field (\autoref{fig:jam_fits1}, \ref{fig:jam_fits2} and \autoref{tab1}). This was not expected. It would have been natural if we had employed e.g. the more general orbit-superposition method \citep[e.g.][]{Cappellari2006}, which is fully described by thousands of parameters. However our models have just six non-linear parameters and one scaling factor $(M/L)_{\rm stars}$. 

Moreover the fits look similarly good even assuming a power-law halo and a constant-anisotropy stellar body (four nonlinear parameters and one scaling factor). The fact that four parameters are able to consistently describe all features of the two-dimensional maps, for the full set of 14 galaxies, then indicates that (i) the (cleaned) SLUGGS data are reliable and (ii) the model assumptions provide a good description of the dynamics and mass distribution of the real galaxies.

An alternative interpretation for the good fits would be that the anisotropy and dark matter variations, and data systematics, conspire to mimic the simple orbital structure and mass distribution assumed by the models. This would be a realistic possibility when studying a single galaxy, given the non full generality of the JAM models, but such a conspiracy is unlikely for such a large and diverse set of galaxies.

\subsection{Isothermal profiles with small scatter to 4\re}

The second and main result of this work is that all 14 modeled fast-rotator ETGs have a nearly-isothermal $\rho_{\rm tot}\propto r^{-2}$ {\em total} density distribution from $\re/10$ out to the median radius of 4\re\ sampled by this study (\autoref{fig:mass_profiles}(a)). The {\em total} mass-density profiles\footnote{Computed from the axisymmetric MGEs as \\
$\rho_{\rm tot}(r)=\sum_{j=1}^M 
\frac{M_j \exp\left[-r^2/(2\sigma_j^2)\right]\, {\rm erf}\left[r\sqrt{1-q_j^2}/(q_j\sigma_j\sqrt{2})\right]}{4\pi\sigma_j^2 r\sqrt{1-q_j^2}}$},
marginalized over all nuisance parameters, are tightly constrained by the data. In the whole range $0.1\re<r<4\re$ the profiles are well described by a power-law $\rho_{\rm tot}\propto r^{-\gamma}$ with the largest average deviation of 11\%. The corresponding average logarithmic slope is $\langle\gamma\rangle=2.19\pm0.03$ for the sample, with an rms scatter of just $\sigma_\gamma=0.11$. We do not detect any significant correlation of the slope with \re, stellar mass, or stellar velocity dispersion. For $0.1\re<r<\re$ the average slope is $\langle\gamma\rangle=2.15\pm0.03$ with $\sigma_\gamma=0.10$, while for $\re<r<4\re$ we find $\langle\gamma\rangle=2.27\pm0.06$ with $\sigma_\gamma=0.23$.

Our inner-profile slope and scatter are in excellent agreement with those $\langle\gamma\rangle=2.08\pm0.03$ with $\sigma_\gamma=0.16$ found around $r\approx\re/2$ using strong lensing \citep{Auger2010}.

\autoref{fig:mass_profiles}(b) shows that the observed trend is consistent with what one would predict for the whole \atl\ sample, for cosmologically-motivated {\em uncontracted} NFW halos. In \autoref{fig:mass_profiles}(c), the stellar profiles are very different from the total ones at the radii we sample: they  have slopes $\rho_{\rm stars}\propto r^{-2}$ around $r\approx\re/2$ (fig.~2 of \citealt{Cappellari2013p15}) but fall off more steeply than $\rho_{\rm stars}\propto r^{-3}$ around $r\approx4\re$.

\autoref{fig:mass_profiles}(d) compares our total profiles with published ones for NGC~0821, NGC~2974, NGC~4494, NGC~4649, and NGC~4697 \citep[from][respectively]{Forestell2010,Weijmans2008,Morganti2013,Das2011,deLorenzi2008}. The $\rho_{\rm tot}(r)$ was derived from the circular velocities $v_c(r)$, assuming spherical symmetry. In 4/5 cases the agreement is excellent, with our {\em statistical} uncertainties enclosing the published profiles over the full radial range. Notable is the tight agreement for NGC~2974, where the $v_c(r)$ was directly measured from a regular HI disk. We believe the disagreement for NGC~4494 may be due to the inclusion in \citet{Morganti2013} models of the strong asymmetry in the SLUGGS data at $r>100\arcsec$. We excluded those data from our fits, but including them would improve the agreement.

\section{Conclusions}

We combine the integral-field stellar kinematics from the \atl\ survey, within $\sim1\re$, with the two-dimensional stellar kinematics from the SLUGGS survey, out to a median radius of about 4\re, and a maximum radius of 2.0--6.2 \re, for a sample of fast rotator early-type galaxies consistent with axisymmetry. We construct the first statistically significant set of detailed axisymmetric dynamical models of the two-dimensional stellar kinematics out to those large radii, where dark matter dominates.

We find that the galaxies' dynamics is well represented by few relatively simple assumptions. The models tightly constrain the {\em total} density profiles, which closely approximates the isothermal form $\rho_{\rm tot}\propto r^{-2}$, from \re/10 out to the median radius of 4\re\ sampled by the data, with remarkably little scatter. The observed total mass distribution is not a generic prediction of $\Lambda$CDM and provides constraints on the models \citep[e.g.][]{Remus2013,Dutton2014}.

Our sample highlights the importance of similar studies on larger samples of galaxies, to provide a much needed benchmark for galaxy formation models. For this, studies like the present one, using DEIMOS on Keck or MUSE on the VLT, can be complemented with models of shallower data, but for much larger samples like MaNGA \citep{Bundy2015}. To be most useful, samples need to include both spiral and early-type galaxies. These should be modeled in a fully homogeneous way by describing the kinematics of both the stars and the gas within the same gravitational potential. Large radio surveys of the neutral HI gas like Apertif \citep{Verheijen2008} can complement MaNGA at the largest radii. The future availability of large kinematic datasets indicates a bright future for the systematic investigation of the mass profiles in galaxies.

\acknowledgements

MC acknowledges support from a Royal Society University Research Fellowship. Supported by National Science Foundation grants AST-0909237 and AST-1211995. DAF thanks the ARC for financial support via DP130100388.

\end{document}